\documentstyle[12pt]{article}
\def\hrul{\rule{15mm}{0mm}}
\def\vrul{\rule{0mm}{6mm}}

\def\A{\kern +0.6ex\lower 0.42ex\hbox{$\scriptstyle\bf\iota$}\kern
-1.20ex a}
\def\E{\kern +0.5ex\lower 0.42ex\hbox{$\scriptstyle\bf\iota$}\kern
-1.10ex e}
\def\be{\begin{eqnarray}}
\def\ee{\end{eqnarray}}

\def\thb{{$t\rightarrow H^+b$}}
\def\thbm{{t\rightarrow H^+b}}
\def\taum{{H^-\rightarrow \tau\bar\nu_\tau}}

\begin{document}
\thispagestyle{empty}
\vspace*{-2cm}
\begin{flushright}
{Alberta Thy-29-93}    \\
{June 1993}           \\
\end{flushright}
\vspace{1cm}
\begin{center}
{\bf \large
Electroweak corrections to decays involving a charged Higgs boson}
\end{center}
\vspace{1cm}
\begin{center}
{Andrzej Czarnecki\footnote{Address after September 1993: Johannes
Gutenberg-Universit\"at, Institut f\"ur Physik, Staudingerweg 7,
6500 Mainz, Germany}}\\
\vspace{.3cm}
{\em Department of Physics, University of Alberta, Edmonton, Canada
T6G 2J1}
\end{center}
\hspace{3in}
\begin{abstract}
We present complete one-loop radiative corrections to the decay
rate of a top quark into a charged Higgs boson and a bottom quark,
and for the decay of a charged Higgs boson into leptons. The results
are discussed in the framework of the two Higgs boson extension of
the Standard Model suggested by supersymmetry. The effect of
electroweak corrections after exclusion of universal
corrections $\Delta r$ is found
to decrease the partial width of the top quark typically by 5\%.
\end{abstract}

\begin{flushleft}
\end{flushleft}

\newpage

\section{Introduction}
Due to the expected large mass of the top quark and its possible
large
Yukawa coupling to Higgs bosons, decays of this particle (once it
is observed, presumably at the Tevatron) can give us an insight into
the
Higgs sector and the mechanism of mass generation. A topic of
particular
importance is the number of Higgs doublets. The supersymmetric
extensions
of the Standard Model, for example, predicts existence of at least
two Higgs doublets. In such scenarios, in addition to the charged
Goldstone
boson of the standard electroweak theory there would be a physical
charged
scalar particle $H^\pm$. Its presence could influence the rate of top
quark decay and even open up a new decay channel.

   If the charged Higgs boson is heavier than the top quark, its
effect
on the decay rate of the top will only be in the virtual corrections
to
the standard process $t\rightarrow W^+ b$.
This has been examined in ref.~\cite{hollik,denner}, and in some
models
the effect was found to be large, of the order of several percent.
On
the other hand, if the decay of the top into the charged Higgs and a
bottom quark is kinematically allowed, it can become the dominant
decay
channel, especially if the ratio of vacuum expectation values of the
two
doublets is such that the Yukawa coupling to the top is not
suppressed.
It is this scenario that is the topic of the
present work.  We examine the effects of first order electroweak
corrections on the width of the decay $t\rightarrow H^+ b$ in the two
Higgs doublet extension of the Standard Model suggested by
supersymmetry
\cite{b272,hunter}.
In this model one of the Higgs fields, $H_1$, is responsible for
giving
masses to down-quarks, and the other one, $H_2$ - to up-quarks.
The ratio of the expectation values of these two fields is denoted by
$\tan\beta=v_2/v_1$. In the present paper we consider the range of
small values of $\tan\beta$, in which  the mass of the bottom quark
can
be safely neglected, which considerably simplifies the calculations.

Radiative corrections to the decay $t\rightarrow H^+ b$
have been subject of several recent publications. The QCD corrections
have been studied in ref.~\cite{liuyao90,liyuan90}. An analysis of
effects of
the mass of the $b$ quark and a comparison of corrections to the main
decay channels $t\rightarrow H^+ b$ and $t\rightarrow W^+ b$ has been
done
in ref.~\cite{sa92b}, where further references on this subject can be
found.

In the electroweak sector the corrections
have been studied only to the order
$O(\alpha m_t^2/m_W^2)$. They have
been calculated in ref.~\cite{csli93} and further analyzed
in \cite{diaz93}. Such corrections would be dominant if the top quark
was much heavier than the $W$ boson. However in view of the expected
mass of the top quark of the order of $(1.5 - 2) m_W$ it is important
to
compute also the remaining corrections not involving the top quark
mass,
as well as the effect of real photon radiation.

This paper is organized as follows: the next section explains
the renormalization
scheme and various kinds of corrections. Section~\ref{sec:cancel}
 discusses cancellation of
infrared and ultraviolet divergences, especially the quadratic ones.
Calculation of virtual corrections to vertices and evaluation of the
bremsstrahlung
are explained in sections~\ref{sec:vertex}
and~\ref{sec:real} . Section~\ref{sec:results} presents final
results;
previously unpublished formulas for renormalization constants are
collected in the Appendix.
\section{Renormalization scheme}
At the tree level the decay rate for \thb\ is obtained from the
Feynman rule
for the $tbH^+$ vertex:
\be
i {e m_t\over \sqrt{2} m_W s_W}\cot\beta\bar b {\rm R} t,
\ee
where we have taken the relevant element of the Kobayashi-Maskawa
matrix to be
equal 1 (and neglected the effect of the $b$ quark mass).
R denotes the right chiral projection operator $(1+\gamma_5)/2$.
We use $m_W$ and
 $m_Z$ as input parameters and define $c_W^2=1-s_W^2=m_W^2/m_Z^2$.
The resulting rate of the decay is:
\be
\Gamma^{(0)}\left( \thbm \right)=
{\alpha m_t^3\over 16m_W^2 s_W^2 }
\cot^2\beta
\left(      1-       {m_{H^+}^2 \over m_t^2 }   \right)^2.
\ee
Electroweak corrections modify the values of parameters in the
vertex: the
coupling constant $e$, masses $m_W$, $m_Z$ and $m_t$,
and the angle $\beta$.
It is also necessary to calculate effects of the real photon
radiation,
virtual corrections to the vertex (triangle diagrams) and the
renormalization
of wave functions of the charged Higgs and of the quarks $t$ and $b$.
On the one loop level there are also
contributions from the mixing of the charged
Higgs with the $W$ boson. Finally, since we are going to work in the
't~Hooft-Feynman gauge, we have to include the mixing between $H^+$
 and the charged Goldstone boson $G^+$. The one loop
correction to the decay rate can be written in the following form:
\be
\Gamma^{(1)}
\left(\thbm\right)=2\Gamma^{(0)}
\left(\thbm\right)\left({\delta e\over e}-{\delta s_W\over s_W}
+\right.{\delta m_t\over m_t}-
{\delta m_W\over m_W}+{\delta \cot\beta\over \cot\beta}
\nonumber \\  \left.
+{1\over 2}\delta^t_{REAL}
+\delta^t_\Delta+{1\over 2}\delta Z^L_b+{1\over 2}\delta Z^R_t+
{1\over 2}\delta Z_H+\delta^t_{MIX}\right).
\label{topcc}
\ee
$\delta^t_{REAL}$, $\delta^t_\Delta$ and $\delta^t_{MIX}$ denote
contributions
of the real photon radiation, triangle diagrams and mixing of $H^+$
with
$W^+$ and with $G^+$ respectively.
For the renormalization of the angle $\beta$ we employ the
prescription
introduced by M\'endez and Pomarol \cite{mend91,mend92}, with a small
 modification. It is assumed that the value of $\beta$ will be
extracted from
 the leptonic decay channel of the charged Higgs boson. Since the
coupling
is proportional to the mass, the dominant decay will be into a $\tau$
lepton
and its neutrino. The renormalization of the angle $\beta$ is fixed
by the
condition that radiative corrections to the vertex $\tau\nu_\tau H$
vanish.
However, the renormalization constant for $\beta$ defined in this way
is
infrared divergent; this problem was not addressed in the original
papers
\cite{mend91,mend92}, because only the fermionic loop corrections
were
discussed there. The infrared divergence could also be removed in the
suitable process of extracting the value of the $\beta$ angle from
the
experimental measurement of the decay width of the charged Higgs
boson.
 For the purpose of the current calculation it is convenient
to include the effect of the real photon radiation in definition of
$\delta\beta$. The one loop correction to the decay rate of the
charged Higgs
into tau and the neutrino  can be written in analogy to the top
decay:
\be\lefteqn{
\Gamma^{(1)}
\left(\taum\right)=2\Gamma^{(0)}
\left(\taum\right)\left({\delta e\over e}-{\delta s_W\over s_W}
+\right.{\delta m_\tau\over m_\tau}-
{\delta m_W\over m_W}}
\nonumber \\&&  \left. -{\delta \cot\beta\over \cot\beta}
+{1\over 2}\delta^\tau_{REAL}
+\delta^\tau_\Delta+{1\over 2}\delta Z^L_\nu
+{1\over 2}\delta Z^R_\tau+
{1\over 2}\delta Z_H+\delta^\tau_{MIX}\right).
\ee
The notation here is analogous to the formula (\ref{topcc}).
Since the coupling of the charged Higgs to leptons is proportional to
$\tan\beta$, the effect of  renormalization of $\beta$ has an
opposite
sign in the two decays under consideration. The reason for this is
that
in both cases we have only one fermion with non-negligible mass, but
they
have opposite values of the weak isospin.

The condition of vanishing of radiative corrections to the
tau channel of the
decay of the charged Higgs allows us to express the renormalization
constant
of the $\beta$ angle in terms of corrections to the $H\tau\nu_\tau$
vertex.
This leads to the following formula for the relative correction to
the rate
\thb:
\be
\lefteqn{\Delta\Gamma\equiv
{\Gamma^{(1)}\left(\thbm\right)\over \Gamma^{(0)}\left(\thbm\right)}}
\nonumber \\
&=&
2\left(2{\delta e\over e}-2{\delta s_W\over s_W}
+\right.{\delta m_\tau\over m_\tau}+{\delta m_t\over
m_t} -{\delta m_W^2\over m_W^2}
\nonumber \\ &&
+{1\over 2}\delta^\tau_{REAL}+{1\over 2}\delta^t_{REAL}
+\delta^\tau_\Delta+\delta^t_\Delta
+{1\over 2}\delta Z^L_\nu
+{1\over 2}\delta Z^R_\tau
\nonumber \\ && \left.
+{1\over 2}\delta Z^L_b+{1\over 2}\delta Z^R_t
+\delta Z_H+\delta^\tau_{MIX}+\delta^t_{MIX}\right).
\ee

As will be seen later, the mixing can be described by one constant
$\delta_{MIX}$ defined so that
\be
\delta^\tau_{MIX}+\delta^t_{MIX}={\cot\beta-\tan\beta\over m_{H^+}^2
-m_W^2}\delta_{MIX}.
\ee

The renormalization of the electroweak parameters is done in the
on-shell
scheme of ref.~\cite{boehm,denner,denn93}.
In particular, for the weak coupling constant $e/s_W$ we have:
\be
{\delta e\over e}-{\delta s_W\over s_W}
&\equiv &\delta Z_e+{\delta m_Z^2\over 2m_Z^2}
-{\delta m_Z^2-\delta m_W^2\over 2(m_Z^2-m_W^2)}
\nonumber\\
&=&
{1\over 2}
\left.{\partial \Sigma^{AA}_T(s)\over \partial s}\right|_{s=0}
-{s_W\over c_W}{\Sigma^{AZ}_T(0)\over m_Z^2}+{\delta m_Z^2\over
2m_Z^2}
-{\delta m_Z^2-\delta m_W^2\over 2(m_Z^2-m_W^2)}.
\ee
This leads to the final formula from which we are going to calculate
the one loop corrections:
\be
\lefteqn{\Delta\Gamma=2\left(2{\delta Z_e}+\right.{\delta m_\tau\over
m_\tau}
+{\delta m_t\over m_t}
-{\delta m_W^2\over m_W^2}+{\delta m_Z^2\over m_Z^2}
-{\delta m_Z^2-\delta m_W^2\over (m_Z^2-m_W^2)}}
\nonumber \\ &&
+{1\over 2}\delta^\tau_{REAL}+{1\over 2}\delta^t_{REAL}
+\delta^\tau_\Delta+\delta^t_\Delta
+{1\over 2}\delta Z^L_\nu
+{1\over 2}\delta Z^R_\tau
\nonumber \\ && \left.
+{1\over 2}\delta Z^L_b+{1\over 2}\delta Z^R_t
+\delta Z_H+{\cot\beta-\tan\beta\over m_{H^+}^2
-m_W^2}\delta_{MIX}
\right).
\label{eq:deltag}
\ee
Many details and explicit formulas for some of the renormalization
constants
can be found in ref.~\cite{hoang92}.
There are no external Higgs particles in
processes described in that reference,
so the wave function renormalization of the charged Higgs boson
and mixing with $W^+$ and Goldstone boson was not included. The
relevant
formulas can be found in the appendix of the present work.

\section{Remarks on  cancellation of divergences}
\label{sec:cancel}
In the calculation of electroweak corrections to decays \thb\ and
$H^-
\rightarrow \tau \bar\nu_\tau$ one encounters three kinds of infinite
quantities: infrared divergences,
 and logarithmic and quadratic ultraviolet divergences.
The infrared divergent integrals result from the radiation of soft
and
collinear photons from external charged particles. They are cancelled
in
the calculation of the total decay rate by wave function
renormalization
constants of the Higgs boson and of fermions, as well as by
corrections to
the Higgs-fermion vertex. For the purpose of the present calculation
the
infrared divergence was regularized by introducing a small mass
$\lambda$
of the photon. All phase space integrals relevant to this problem
have
been listed in ref.~\cite{denn93}.

The ultraviolet divergent integrals are regularized dimensionally.
In this scheme, the quadratic divergences show up as poles at
number of dimensions $n=2$.
They originate from tadpole diagrams and from the
fermionic loop contribution to charged Higgs - Goldstone boson
mixing.
Some individual non-tadpole diagrams in boson self energies  also
contain
quadratic divergences, but the relevant sums of diagrams are free
from
them (in the
't~Hooft-Feynman gauge), just like in the Standard Model
\cite{velt81}.
Goldstone bosons are absent in the unitary gauge and there all the
tadpole
contributions cancel out. The problem is more delicate in the
't~Hooft-Feynman gauge, in which the present calculation is
done\footnote{A discussion of tadpole diagrams with a fermion loop
can be found in ref.~\cite{cap91} which also contains further
references.}.

The different types of tadpole diagrams in the two Higgs doublet
model are
shown in figure~(\ref{fig:tadpol}). The external particle can be one
of
the CP even neutral Higgs bosons, $H^0$ or $h^0$. These diagrams
contribute
to mass renormalization
 of external fermions, to $\delta m_W$ and $\delta m_Z$, and to the
mixing
between the Higgs boson and Goldstone and $W$ bosons. The quadratic
divergence from the fermionic loop in figure~(\ref{fig:tadpol}b)
cancels the
one from the fermionic contribution
to the Higgs-Goldstone mixing shown in figure~(\ref{fig:mixss}a). The
sum of
contributions of the
remaining, bosonic  tadpole diagrams, is free from
quadratic divergences.
The logarithmic divergences of tadpole diagrams are cancelled by loop
diagrams of Higgs-Goldstone
boson mixing depicted in figures~(\ref{fig:mixtad}a)
and~(\ref{fig:mixss}b,c).
 The sum of bosonic loops of Higgs-$W$ boson mixing is finite.

\section{Vertex corrections}
\label{sec:vertex}
Electroweak
corrections to vertices are of two kinds: there are modifications of
the
values of parameters determining the strength of the coupling and
relations
among them, and triangle diagrams. It is this second type which will
be
considered in this section. The basic types of triangle
diagrams contributing to
decays of the $t$ quark and the charged Higgs boson are depicted in
figures
(\ref{fig:topv}) and  (\ref{fig:tauv}). Since the number of diagrams
of
is fairly large it is most convenient to employ the method of
standard
matrix elements (see ref.~\cite{denn93} for a review and further
references).
The principle of this method is to calculate coefficients in a
representation
of an invariant matrix element in form of a sum over certain standard
tensors, which depend only on the Lorentz structure of the process.
In
particular, in the case of scalar-fermion interaction, there are only
two
standard matrix elements:
\be
{\cal M}^L&=&\bar u(p) L u(q), \nonumber \\
{\cal M}^R&=&\bar u(p) R u(q),
\ee
where $L=(1-\gamma_5)/2$.
Born amplitude of  the decay of the Higgs boson into leptons is
proportional to ${\cal M}^L$, and since on the level of one-loop
corrections we need to compute  only the interference  of  triangle
and tree
diagrams, it is sufficient to evaluate only the ${\cal M}^L$
component
of the triangle diagrams.
\begin{table}
\begin{center}
\begin{tabular}{ccccl}
      \hline      \hline
Diagram &\multicolumn{3}{l}{Particle assignments}& \\
(Figure No.) &                X & Y & Z & Residuum \\
\hline
\ref{fig:topv}(a)
&$ t $&$ H^+ $ &$ H^0$      &   0  \\
\ref{fig:topv}(a)
&$ t $&$ H^+ $  &$ h^0 $    &   0    \\
\ref{fig:topv}(a)
&$ t $&$ G^+ $ &$ A^0 $     &   0      \\
\ref{fig:topv}(b)
&$ t $&$ H^+ $ &$ \gamma $  &   2/3      \\
\ref{fig:topv}(b)
&$ t $&$ H^+ $ &$ Z $       &  $-1/3 + s_W^2/(3 c_W^2) $
\\
\ref{fig:topv}(c)
&$ t $&$ W^+ $   &$ H^0 $    &
   $-\sin\alpha\sin(\beta-\alpha)/(4s_W^2\cos\beta)$ \\
\ref{fig:topv}(c)
&$ t $&$ W^+ $  &$ h^0 $    &
   $\cos\alpha\cos(\beta-\alpha)/(4s_W^2\cos\beta)     $       \\
\ref{fig:topv}(c)
&$ t $&$ W^+ $  &$ A^0 $     &$  1/(4s_W^2)     $      \\
\ref{fig:topv}(c)
&$ b $&$ \gamma $ &$ H^+ $   & $ 1/3                $    \\
\ref{fig:topv}(c)
&$ b $&$ Z $ &$ H^+ $        & $
(2s_W^2-3)(s_W^2-c_W^2)/(12s_W^2c_W^2)$   \\
\ref{fig:topv}(d)
&$ \gamma $&$ b $ &$ t $     &$  -8/9              $         \\
\ref{fig:topv}(d)
&$ Z $&$ b $   &$ t $        & $ (12-8 s_W^2)/(9 c_W^2) $         \\
\hline
\ref{fig:tauv}(a)
&$\gamma$ &$ \tau $& $H^-$   &$ 1                       $ \\
\ref{fig:tauv}(a)
&$Z$ & $\tau$ & $H^-$        &$ (s_W^2-c_W^2)/(2c_W^2)$\\
\ref{fig:tauv}(b)
&$W$ & $\tau$ & $H^0$        &$ \cos\alpha
\sin(\beta-\alpha)/(4s_W^2\sin\beta) $ \\
\ref{fig:tauv}(b)
&$W$ & $\tau$ & $h^0 $       &$ \sin\alpha
\cos(\beta-\alpha)/(4s_W^2\sin\beta)$\\
\ref{fig:tauv}(b)
&$W$ & $\tau$ & $A^0$        &$ 1/(4s_W^2)            $ \\
\ref{fig:tauv}(b)
&$Z$ & $\nu_\tau$ & $H^-$      &$ (c_W^2-s_W^2)/(2s_Wc_W)^2$\\
\ref{fig:tauv}(c)
&$Z$ & $\nu_\tau$ & $\tau$   &$ 2/c_W^2 $\\
\hline\hline
\end{tabular}
\end{center}
\caption{Particle contents of triangle diagrams. The last column
shows the coefficient of ${\alpha\over 4\pi}{2\over 4-n}$ in
$\delta^t_\Delta$
and $\delta^\tau_\Delta$.}
\label{tab:triang}
\end{table}
Analogously, in the case of the top quark
decay, we need the ${\cal M}^R$ part only. The resulting formulas are
quite space consuming and will not be shown here, because in contrast
to
the two point functions their applicability in other contexts is
rather
limited. However, in Table~\ref{tab:triang} we list concrete
  particle assignments to the general diagrams of figures
(\ref{fig:topv}) and  (\ref{fig:tauv}) together with explicit
expressions of their ultraviolet
divergent parts.

Complete analytic formulas are obtained using {\it FeynArts}
 (also used to illustrate the present paper) and {\it
 FeynCalc}~\cite{feynarts,feyncalc}. Fortran output of
these programs is evaluated using the library {\it FF} \cite{ff}.

\newpage
\section{Real photon radiation}\label{sec:real}
Triangle diagrams discussed in the previous section are infrared
divergent
due to exchange of soft photons. These divergences are cancelled by
bremsstrahlung processes depicted in figures (\ref{fig:topreal}) and
(\ref{fig:taureal}). These diagrams can be easily evaluated in terms
of
phase space integrals listed in ref.~\cite{denn93}. We give here as
an
example an explicit formula for the width of the process
$H^-\rightarrow
\tau\bar\nu_\tau\gamma$:
\be
\Gamma\left(H^-\rightarrow\tau\bar\nu_\tau\gamma\right)
={e^4m_\tau^2\tan^2\beta\over 2^7\pi^3s_W^2m_{H^-}m_W^2}
\left(|{\cal A}|^2+|{\cal B}|^2+{\cal A}^*{\cal B}+{\cal B}^*{\cal
A}\right),
\ee
where ${\cal A}$ and ${\cal B}$ denote the amplitudes corresponding
to
 diagrams in figure (\ref{fig:taureal}), for which we have:
\be
|{\cal A}|^2&=&
 4m_{H^-}^2 (m_\tau^2- m_{H^-}^2)
 I_{00} + 2( m_\tau^2   - 3 m_{H^-}^2 )I_0 - 2 I,
\nonumber \\
|{\cal B}|^2&=&
      4m_\tau^2  (m_\tau^2  -
m_{H^-}^2) I_{11} + 4 m_\tau^2 I_1
         - 2 I - 2 I_1^0,
\nonumber \\
{\cal A}^* {\cal B}+{\cal B}^* {\cal A}&=&
      4( m_\tau^4  - m_{H^-}^4) I_{01}
+2( m_\tau^2+ m_{H^-}^2 ) I_0
- 4 m_{H^-}^2 I_1+2I.
\ee
Integrals $I$ are taken from ref.~\cite{denn93}, where explicit
expressions
can be found. Here we only quote the definition:
\be
I^{j_1,...,j_m}_{i_1,...,i_n}={1\over \pi^2}
\int {d^3p_1\over 2p_{10}}{d^3p_2\over 2p_{20}}{d^3q\over 2q_{0}}
\delta^{(4)}\left( p_0-p_1-p_2-q \right)
{(\pm 2qp_{j_1})...(\pm 2qp_{j_m})\over(\pm 2qp_{i_1})...(\pm
2qp_{i_n})},
\ee
where we $q$, $p_0$, $p_1$ and $p_2$ denote momenta of the photon,
Higgs boson, tau and neutrino respectively, and the signs should be
chosen
in the following way: minus sign if $i_k$ or $j_k$ is zero, plus in
all other
cases. Functions  $I_{00}$, $I_{01}$ and $I_{11}$
contain infrared divergences,
regularized by introducing a small mass of the photon $\lambda$. If
the mass
of neutral particle in the final state is small, the representation
of these functions given in \cite{denn93} becomes numerically
unstable and
it is more convenient to use the corresponding formulas from
ref.~\cite{cza90}.

\section{Results and discussion}\label{sec:results}
Following ref.~\cite{denner,hoang92}, the electroweak correction can
be
expressed by comparing the one-loop decay width to the Born rate
parameterized by Fermi coupling constant $G_F$ instead of the fine
structure
constant $\alpha$:
\be
\Gamma^{(0)}(G_F)={\Gamma^{(0)}(\alpha)\over 1-\Delta r},
\ee
where $\Delta r$ denotes radiative corrections to the muon decay,
from which
Fermi constant is determined.
 Such representation has the advantage of including
large corrections due to fermion loops in the Born rate.
In the present renormalization scheme, based on the condition of
vanishing of radiative corrections to the $H^+\tau\nu_\tau$ vertex,
the effect
of coupling constant renormalization is doubled (see
equation~\ref{eq:deltag}), and one ought to subtract $2\Delta r$ in
order
to cancel the fermion loop contribution from universal corrections.
This is due to the fact that in order to avoid the artificially large
corrections one has to parameterize {\it both} decay rates
$\Gamma^0\left(
H^\pm\rightarrow \tau\nu_\tau\right)$ and $\Gamma^0\left(t\rightarrow
H^+b
\right)$ by $G_F$. At this point our analysis differs from
ref.~\cite{csli93}.
For moderate values of $\tan\beta>1$ the corrections consist
typically of
-4\% bosonic contributions and +7\% from fermion loops. This last
part is
cancelled by subtraction of $\Delta r$, so that the fermionic
contribution
to the corrections becomes slightly negative. This can be seen in
figure
(\ref{fig:coranat}).

Numerical evaluation of corrections to the decay width  $\Gamma^{(0)}
(G_F)$ proceeds in the following way. The set of input parameters
consists
of $m_Z$, $G_F$, $\alpha$, masses of fermions and CKM matrix
elements;
values of them are taken from a recent review \cite{denn93}. All the
numerical results are presented for mass of the top quark equal 140
GeV.
In addition
we need two parameters of the Higgs sector: we choose angle $\beta$
and
mass of the charged Higgs boson. Masses of the remaining Higgs
particles and
angle $\alpha$ are found using the formulas of ref.~\cite{hunter}.
Mass of
the $W$ boson is found by solving a nonlinear equation \cite{denner}:
\be
m_W^2\left(1-{m_W^2\over m_Z^2}\right)={\pi\alpha\over \sqrt{2} G_F}
{1\over 1-\Delta r}.
\ee
Finally, using this value of $m_W$, we find $\Delta r$ and
$\Delta\Gamma$.
The resulting corrections
$\overline{\Delta\Gamma}=\Delta\Gamma-2\Delta r$
are plotted as a function of mass of the charged Higgs boson in
figure
(\ref{fig:cormhp}) and as function of $\tan\beta$ in
(\ref{fig:cortanb}).

Similarly to the case of the decay $t\rightarrow W^+b$~\cite{denner},
the corrections become large when mass of the lighter CP even neutral
Higgs boson $h^0$ is small. In particular,
they  diverge at the point $\tan\beta=1$ where $m_{h^0}=0$. This
divergence
should be cancelled by adding width of the decay $t\rightarrow
H^+bh^0$,
just like the infrared divergence due to virtual photon exchange is
cancelled
by the real photon radiation. As $\tan\beta$ becomes larger (or
smaller)
than 1,
mass of $h^0$
increases, and at the point where it reaches $m_{H^+}-m_{W^+}$,
amplitudes
of both decays $H^\pm\rightarrow \tau\nu_\tau$ and $t\rightarrow
H^+b$
have singularities which show up as discontinuities of the derivative
of the one-loop decay rate and can be noticed on the diagrams; the
value
of $\tan\beta$ where this happens is close to 1 for light $H^+$, and
gets further away as $H^+$ becomes heavier. The corresponding cusps
on the
diagrams are easier to recognize for $\tan\beta>1$, but they are
present
also in the region of $\tan\beta<1$.

\section*{Acknowledgment}
The author thanks Professor A.N. Kamal for many helpful discussions
and
for clarifying the problem of $W$-Higgs-Goldstone boson mixing, and
Dr.~J.~Pinfold for the opportunity to use the computing facilities of
the
Subatomic Research Center of the University of Alberta.
It is a pleasure to acknowledge helpful correspondence with
Dr.~G.J.~van~Oldenborgh about his library FF.
Many thanks go to
B.~Darian, S.~Droz and J.~MacKinnon for sharing their expertise in
computing.

This research was supported by Killam Foundation and a Dissertation
Fellowship
 of the University of Alberta, and by a grant to Professor A.N. Kamal
from
the Natural Sciences and Engineering Research Council of Canada.

\appendix
\section{Renormalization constants}
In this Appendix we list those of renormalization constants of the 2
Higgs
doublet model which have not been published so far. We first give
expressions
for the wave function renormalization of the charged Higgs boson and
then
analyze various contributions to the mixing of Higgs boson with $W$
and
Goldstone boson $\delta_{MIX}$. The results are given in terms of
standard
Passarino-Veltman integrals \cite{pave}, using the conventions of
ref.~\cite{denn93,hoang92}, where many useful properties of these
functions
have been collected.

The wave function renormalization constant of the charged Higgs boson
gets
contributions from diagrams with  fermion, scalar and vector-scalar
loops.
To make the formula more compact it is convenient to introduce the
notation:
\be
\lambda\left(m_i,m_j,m_k\right)\equiv m_i^4+m_j^4+m_k^4
         -2m^2_im^2_j-2m^2_im^2_k-2m^2_jm^2_k.
\ee
The bosonic contributions to the renormalization constant $\delta
Z_H$ is:
\be
\lefteqn{\delta Z_H^{\rm bos} ={\alpha\over
4\pi}\left\{\sum_{H=H^0,h^0,A^0}
\left\{ {1\over 4m_W^2s_W^2} \right.\right.}
\nonumber\\&&
\left( \delta_{HH^0}\sin^2(\beta-\alpha)+
\delta_{Hh^0}\cos^2(\beta-\alpha)+\delta_{HA^0}\right)
\nonumber\\ && \left.
\left[ 2m_W^2B_0 \left(m_{H^+}^2,m_H,m_W\right)
-\lambda\left(m_{H^+},m_{H},m_W\right)
B_0^\prime\left(m_{H^+}^2,m_H,m_W\right)
\right] \vrul \right\}
\nonumber\\ &&
+{(s_W^2-c_W^2)^2\over 4s_W^2c_W^2} \left[
   2B_0\left(m_{H^+}^2,m_{H^+},m_Z\right)
\right. \nonumber\\ && \left.
\hrul\hrul +\left( 4m_{H^+}^2-m_Z^2\right)
B_0^\prime \left(m_{H^+}^2,m_{H^+},m_Z\right)\right]
\nonumber\\ &&
+2 B_0\left(m_{H^+}^2,m_{H^+},\lambda \right)
+4m_{H^+}^2B_0^\prime \left(m_{H^+}^2,m_{H^+},\lambda \right)
\nonumber\\
&&-{m_W^2\over s_W^2}\left\{ \left[\cos(\beta-\alpha)-{\cos 2\beta
\cos(\beta+\alpha)\over 2 c_W^2} \right]^2
B_0^\prime \left(m_{H^+}^2,m_{H^+},m_{H^0}\right) \right.
\nonumber\\
&& \left.\left.
\qquad+\left[\sin(\beta-\alpha)+{\cos 2\beta \sin(\beta+\alpha)\over
2 c_W^2}
\right]^2
 B_0^\prime \left(m_{H^+}^2,m_{H^+},m_{h^0}\right)\right\}\right\}.
\ee
The contribution of one generation of  quarks is:
\be
\delta Z_H^{\rm q} &=&{\alpha\over 4\pi}{N_C\over 2s_W^2m_W^2}
\left\{ -
\left(  m_d^2\tan^2\beta+m_u^2\cot^2\beta  \right)
B_0\left(m_{H^+}^2,m_d,m_u\right)
\right.
\nonumber\\&&
+\left[\left(
m_d^2\tan^2\beta+m_u^2\cot^2\beta\right)
\left(m_d^2+m_u^2-m_{H^+}^2\right)+4m_d^2 m_u^2 \right]
\nonumber\\&& \left.
\hrul\hrul\hrul B_0^\prime\left(m_{H^+}^2,m_d,m_u\right) \right\}.
 \label{eq:zhq}
\ee
Finally, the contribution of a lepton-neutrino pair is obtained from
the
formula  (\ref{eq:zhq}) by taking $N_C=1$, $m_u=0$ and using:
\be
{d\over ds}B_0\left(s,0,m\right)=
{1\over m^2-s}
\left\{
-{m^2\over s}
\left[B_0\left(s,0,m\right)-B_0\left(0,0,m\right)\right] +1\right\}.
\ee
The result is:
\be
\delta Z_H^{\rm l} &=&{\alpha\over 4\pi}{m^2\tan^2\beta\over 2s_W^2
m_W^2}\left\{
-{m^2\over m_{H^+}^2}\left[B_0\left(m_{H^+}^2,0,m\right)
   -B_0\left(0,0,m\right)\right]\right.
 \nonumber\\
&&\left.  +1-B_0\left(m_{H^+}^2,0,m\right)
\right\}.
\ee

The contribution of bosons to mixing can be represented by the
following
formula:
\be
\lefteqn{\delta_{MIX}^{\rm bos}
={\alpha\over 4\pi s_W^2}\sum_{H=H^0,h^0}\left[
{\sin(\beta-\alpha)\cos(\beta-\alpha)
\over 4}\right.  \left(\delta_{H,H^0}-\delta_{H,h^0}  \right)}
\nonumber\\
&&\cdot\left\{ {(m_H^2-m_W^2)^2\over m_{H^+}^2} \left(
B_0\left( m_{H^+}^2,m_H,m_W \right)-
B_0\left( 0,m_H,m_W \right) \right)
\right. \nonumber\\    &&                          \left.
+\left(2m_{H^+}^2+m_H^2-3m_W^2  \right)
B_0\left( m_{H^+}^2,m_H,m_W \right) \vrul \right\}
\nonumber\\&&
+{m_W^2\over 2m_{H^+}^2} \left\{
\delta_{H,H^0}\sin(\beta-\alpha)
\left( \cos(\beta-\alpha)-{\cos 2\beta\cos(\beta+\alpha)\over 2c_W^2
} \right)
\right. \nonumber\\&& \left.
\hrul - \delta_{H,h^0}\cos(\beta-\alpha)
\left(\sin(\beta-\alpha)+{\cos 2\beta\sin(\beta+\alpha)\over 2c_W^2
}\right)
\right\}
\nonumber\\&&
\cdot\left(m_H^2- m_{H^+}^2\right)
\left[\left(1- {m_{H^+}^2\over m_W^2}\right)
B_0\left( m_{H^+}^2,m_{H^+},m_H \right)-
B_0\left( 0,m_{H^+},m_H\right)  \right]
\nonumber\\ &&
+{\cos 2\beta\over 4c_W^2}
\left(\sin(\beta-\alpha)\cos(\beta+\alpha)\delta_{H,H^0}
+\sin(\beta+\alpha)\cos(\beta-\alpha)\delta_{H,h^0}  \right)
\nonumber\\ &&                         \left.
\hrul \cdot\left( m_{H^+}^2-m_H^2\right)
B_0\left( m_{H^+}^2,m_H,m_W\right) \vrul \right]
\nonumber\\ &&
-{\alpha\over 4\pi}{1\over 8s_W^2c_W^2}\left\{
\sin 2\beta\cos 2\beta \left[
4A(m_W)-4A(m_{H^+})+A(m_Z)-A(m_{A^0})\right]
\right. \nonumber\\ && \left.
\hrul
+\left(c_W^2 \sin 2\alpha \cos 2\beta +s_W^2 \cos 2\alpha \sin 2\beta
\right)\left[A(m_{H^0})-A(m_{h^0})\right]\right\}
\nonumber\\ &&
+{g\over
2m_W}\left[\left(m_W^2-m_{H^+}^2+m_{H^0}^2\right)\sin(\beta-\alpha)
t_1
\right. \nonumber\\ && \left.
\hrul -
\left(m_W^2-m_{H^+}^2+m_{h^0}^2\right)\cos(\beta-\alpha)t_2\right].
\ee
The last two lines in the above formula represent
contributions of tadpole diagrams.
Formulas for  fermion loops are given below for $H-G$ and $H-W$
mixing
separately:
\be
\lefteqn{\delta_{MIX}^{HG}= -{\alpha\over 4\pi}{N_C\over 2s_W^2m_W^2}
\left\{\left(-m_d^2\tan\beta+m_u^2\cot\beta\right) \right. }
\nonumber \\&&
\cdot\left[ (m_d^2+m_u^2-m_{H^+}^2) B_0\left(m_{H^+}^2,m_u,m_d\right)
+A(m_u)+A(m_d)\right]
\nonumber \\&& \left.
+2m_u^2m_d^2(\tan\beta-\cot\beta )
B_0\left(m_{H^+}^2,m_u,m_d\right)\right\},
\nonumber \\
\lefteqn{\delta_{MIX}^{HW}= -{\alpha\over 4\pi}{N_C\over s_W^2}
\left[\left(m_d^2\tan\beta+m_u^2\cot\beta\right)
B_1\left(m_{H^+}^2,m_u,m_d\right)
\right. }\nonumber \\&& \left.
+m_u^2\cot\beta
B_0\left(m_{H^+}^2,m_u,m_d\right)\right].
\ee

of
of
line),

\section*{Figure captions}
\begin{enumerate}
\item
Real photon corrections to the decay of $t$ quark
\label{fig:topreal}

\item
Real photon corrections to the charged Higgs boson decay
\label{fig:taureal}

\item
Vertex corrections to the decay of $t$ quark
\label{fig:topv}

\item
Vertex corrections to the charged Higgs boson decay
\label{fig:tauv}

\item
Momentum independent contributions to mixing
\label{fig:mixtad}

\item
Mixing between the charged Higgs and the $W$ boson
\label{fig:mixsw}

\item
Mixing between the charged Higgs and the Goldstone boson
\label{fig:mixss}

\item
Types of tadpole diagrams in 2HDM
\label{fig:tadpol}

\item
Corrections $\overline{\Delta\Gamma}$ plotted as a function of
$\tan\beta$ for two different values of $m_{H^+}$: $m_{H^+}=90$ GeV
(solid line) and $m_{H^+}=120$ GeV (dashed)
\label{fig:cortanb}

\item
Corrections $\overline{\Delta\Gamma}$ plotted as a function of
$m_{H^+}$ for various values of $\tan\beta$: $\tan\beta=0.5$ (solid
line),
$\tan\beta=1.5$ (long dash) and $\tan\beta=5$ (short dash)
\label{fig:cormhp}

\item
Bosonic contributions to corrections
${\Delta\Gamma}$ (solid line) and  the fermionic contributions from
which twice the value of universal corrections
${\Delta r}$ was subtracted (dashed). Plotted as a function of
$m_{H^+}$ for  $\tan\beta=1.5$
\label{fig:coranat}
\end{enumerate}


\begin{thebibliography}{10}

\bibitem{hollik}
B.~Grz{\A}dkowski and W.~Hollik.
\newblock Radiative corrections to the top quark width within
  two-{Higgs}-doublet models.
\newblock {\em Nucl. Phys.}, B384:101--112, 1992.

\bibitem{denner}
A.~Denner and A.H. Hoang.
\newblock The top decay {$t\rightarrow bW$} in the two {Higgs}
doublet model.
\newblock Karlsruhe preprint TTP 92-24, 1992.

\bibitem{b272}
J.F. Gunion and H.E. Haber.
\newblock {Higgs} bosons in supersymmetric models ({I}).
\newblock {\em Nucl. Phys.}, B272:1--76, 1986.
\newblock Errata: bulletin board hep-ph@xxx.lanl.gov/9301205.

\bibitem{hunter}
J.~F. Gunion, H.~E. Haber, G.~Kane, and S.~Dawson.
\newblock {\em The {Higgs} Hunter's Guide}.
\newblock Addison-Wesley, Redwood City, 1990.
\newblock Errata: bulletin board hep-ph@xxx.lanl.gov/9302272.

\bibitem{liuyao90}
J.~Liu and Y.~P. Yao.
\newblock {QCD} correction to heavy top exclusive decays.
\newblock preprint UM-TH-90-09, 1990.

\bibitem{liyuan90}
C.S. Li and T.C. Yuan.
\newblock {QCD} correction to charged {Higgs} decay of the top quark.
\newblock {\em Phys. Rev.}, D42:3088, 1990.
\newblock Erratum: ibid. D47:2156, 1993.

\bibitem{sa92b}
A.~Czarnecki and S.~Davidson.
\newblock {QCD} corrections to the charged {Higgs} decay of a heavy
quark.
\newblock Alberta preprint Thy-34-92, submitted for publication in
Physical
  Review D, 1992.

\bibitem{csli93}
C.S. Li, B.Q. Hu, and J.M. Yang.
\newblock Electroweak radiative corrections to $t\rightarrow {H^+} b$
for a
  heavy top quark.
\newblock {\em Phys. Rev.}, D47:2865--2871, 1993.

\bibitem{diaz93}
M.A.~D\'\i az.
\newblock Top quark and charged {H}iggs: linked by radiative
corrections.
\newblock Preprint VAND-TH-93-5.

\bibitem{mend91}
A.~M\'endez and A.~Pomarol.
\newblock t-quark loop corrections to the charged {H}iggs boson
hadronic width.
\newblock {\em Phys. Lett.}, B265:177--181, 1991.

\bibitem{mend92}
A.~M\'endez and A.~Pomarol.
\newblock t-quark loop corrections to neutral {H}iggs couplings in
the
  two-{H}iggs-doublet model.
\newblock {\em Phys. Lett.}, B279:98--105, 1991.

\bibitem{boehm}
M.~{B\"ohm} and A.~Denner.
\newblock Radiative corrections in the electroweak {S}tandard
{M}odel.
\newblock In A.~Perez and Roberto Huerta, editors, {\em Proc. of the
Workshop
  on High Energy Phenomenology, Mexico City}, pages 1--113, River
Edge, N.J.,
  1991. World Scientific.

\bibitem{denn93}
A.~Denner.
\newblock Techniques for the calculation of electroweak radiative
corrections
  at the one-loop level and results for {W}-physics at {LEP200}.
\newblock {\em Fortschr.~Phys.}, 41, 1993.
\newblock In press.

\bibitem{hoang92}
A.~Hoang.
\newblock Tau- und {T}op-{Zerf\"alle} im
{Z}wei-{H}iggs-{D}ublett-{M}odell.
\newblock Master's thesis, Karlsruhe University, 1992.
\newblock In German.

\bibitem{velt81}
M.~Veltman.
\newblock The infrared-ultraviolet connection.
\newblock {\em Acta Phys. Pol.}, B12:437--457, 1981.

\bibitem{cap91}
M.~Capdequi Peyran{\`e}re, H.E. Haber, and P.~Irulegui.
\newblock {$H^\pm\rightarrow W^\pm\gamma$} and {$H^\pm\rightarrow
W^\pm Z$} in
  two-{H}iggs-doublet models: large-fermion-mass limit.
\newblock {\em Phys. Rev.}, D44:191--201, 1991.

\bibitem{feynarts}
J.~K{\"u}hlbeck, M.~B{\"o}hm, and A.~Denner.
\newblock Feyn {A}rts - computer-algebraic generation of {F}eynman
graphs and
  amplitudes.
\newblock {\em Comp. Phys. Comm.}, 60:165--180, 1990.

\bibitem{feyncalc}
R.~Mertig, M.~B{\"o}hm, and A.~Denner.
\newblock Feyn {C}alc - computer-algebraic calculation of {F}eynman
diagrams.
\newblock {\em Comp. Phys. Comm.}, 64:345--359, 1991.

\bibitem{ff}
G.J. van Oldenborgh.
\newblock {FF} - a package to evaluate one-loop {F}eynman diagrams.
\newblock {\em Comp. Phys. Comm.}, 48:1, 1991.

\bibitem{cza90}
A.~Czarnecki.
\newblock {QCD} corrections to the decay {$t\rightarrow Wb$} in
dimensional
  regularization.
\newblock {\em Phys. Lett.}, B252:467--470, 1990.

\bibitem{pave}
G.~Passarino and M.~Veltman.
\newblock One loop corretions for $e^+e^-$ annihilation into
$\mu^+\mu^-$ in
  the {W}einberg model.
\newblock {\em Nucl. Phys.}, B160:151, 1979.

\end{thebibliography}
\end{document}